\documentclass{bioinfo}

\pdfoutput=1
\usepackage{pdfpages}
\usepackage{xcolor}

\usepackage{multirow}
\usepackage{subfigure}

\newcommand{\xuan}[1]{{\color{black}#1}}

\copyrightyear{2015} \pubyear{2015}

\access{Advance Access Publication Date: Day Month Year}
\appnotes{Manuscript Category}

\begin{document}
	\firstpage{1}
	
	\subtitle{Subject Section}
	
	\title[Cross-type Biomedical Named Entity Recognition with Deep Multi-Task Learning]{Cross-type Biomedical Named Entity Recognition with Deep Multi-Task Learning}
	\author[Sample \textit{et~al}.]{Xuan Wang\,$^{\text{\sfb 1,}*}$, Yu Zhang\,$^{\text{\sfb 1}}$, Xiang Ren\,$^{\text{\sfb 2,}*}$, Yuhao Zhang\,$^{\text{\sfb 3}}$, Marinka Zitnik\,$^{\text{\sfb 4}}$, Jingbo Shang\,$^{\text{\sfb 1}}$, Curtis Langlotz\,$^{\text{\sfb 3}}$ and Jiawei Han\,$^{\text{\sfb 1}}$}
	\address{$^{\text{\sf 1}}$Department of Computer Science, University of Illinois at Urbana-Champaign, Urbana, IL 61801, USA, \\
		$^{\text{\sf 2}}$Department of Computer Science, University of Southern California, Los Angeles, CA 90089, USA, \\
		$^{\text{\sf 3}}$School of Medicine, Stanford University, Stanford, CA 94305, USA, and\\
		$^{\text{\sf 4}}$Department of Computer Science, Stanford University, Stanford, CA 94305, USA}
	
	\corresp{$^\ast$To whom correspondence should be addressed.}
	
	\history{Received on XXXXX; revised on XXXXX; accepted on XXXXX}
	
	\editor{Associate Editor: XXXXXXX}
	
	\abstract{\textbf{Motivation:} State-of-the-art biomedical named entity recognition (BioNER) systems often require handcrafted features specific to each entity type, such as genes, chemicals and diseases. 
		Although recent studies explored using neural network models for BioNER to free experts from manual feature engineering, the performance remains limited by the available training data for each entity type. \\
		\textbf{Results:} We propose a multi-task learning framework for BioNER to collectively use the training data of different types of entities and improve the performance on each of them. In experiments on 15 benchmark BioNER datasets, our multi-task model achieves substantially better performance compared with state-of-the-art BioNER systems and baseline neural sequence labeling models. Further analysis shows that the large performance gains come from sharing character- and word-level information among relevant biomedical entities across differently labeled corpora. \\
		\textbf{Availability:} Our source code is available at \href{https://github.com/yuzhimanhua/lm-lstm-crf}{https://github.com/yuzhimanhua/lm-lstm-crf}.\\
		\textbf{Contact:} \href{xwang174@illinois.edu}{xwang174@illinois.edu}, \href{xiangren@usc.edu}{xiangren@usc.edu}. \\
		\textbf{Supplementary information:} Supplementary data are available at \textit{Bioinformatics} online.
	}
	
	\maketitle
	
	\section{Introduction}
	Biomedical named entity recognition (BioNER) is one of the most fundamental task in biomedical text mining that aims to automatically recognize and classify biomedical entities (e.g., genes, proteins, chemicals and diseases) from text. BioNER can be used to identify new gene names from text (\citealp{smith2008overview}). It also serves as a primitive step of many downstream applications, such as relation extraction (\citealp{cokol2005emergent}) and knowledge base completion (\citealp{szklarczyk2017string, wei2013pubtator, xie2013mircancer, szklarczyk2015stitch}).
	
	BioNER is typically formulated as a sequence labeling problem whose goal is to assign a label to each word in a sentence. State-of-the-art BioNER systems often require handcrafted features (e.g., capitalization, prefix and suffix) to be specifically designed for each entity type (\citealp{ando2007biocreative, leaman2016taggerone, guodong2004exploring, lu2015chemdner}). This feature generation process takes the majority of time and cost in developing a BioNER system (\citealp{leser2005makes}), and leads to highly specialized systems that cannot be directly used to recognize new types of entities. The accuracy of the resulting BioNER tools remains a limiting factor in the performance of biomedical text mining pipelines (\citealp{huang2015community}).
	
	Recent NER studies consider neural network models to automatically generate quality features (\citealp{chiu2016named, ma2016end, lample2016neural, 2017arXiv170904109L}). \citeauthor{korhonen2017neural} took each word token and its surrounding context words as input into a convolutional neural network (CNN). \citeauthor{habibi2017deep} adopted the model from \citeauthor{lample2016neural} and used word embeddings as input into a bidirectional long short-term memory-conditional random field (BiLSTM-CRF) model. These neural network models free experts from manual feature engineering. However, these models have millions of parameters and require very large datasets to reliably estimate the parameters. This poses a major challenge for biomedicine, where datasets of this scale are expensive and slow to create and thus neural network models cannot realize their potential performance to the fullest (\citealp{camacho2018next}). Although neural network models can outperform traditional sequence labeling models (e.g., CRF models (\citealp{lafferty2001conditional})), they are still outperfomed by handcrafted feature-based systems in multiple domains (\citealp{korhonen2017neural}).
	
	One direction to address the above challenge is to use the labeled data of different entity types to augment the training signals for each of them, as information like word semantics and grammatical structure may be shared across different datasets. However, simply combining all datasets and \xuan{training} one single model over multiple entity types can introduce many false negatives because each dataset is typically specifically annotated for one or only a few entity types. For example, combining dataset A for gene \xuan{recognition} and dataset B for chemical recognition will result in missing chemical entity labels in dataset A and missing gene entity labels in dataset B. Multi-task learning (MTL) (\citealp{collobert2008unified, sogaard2016deep}) offers a solution to this issue by collectively training a model on several related tasks, so that each task benefits model learning in other tasks without introducing additional errors. 
	MTL has been successfully applied in natural language processing (\citealp{collobert2008unified}), speech recognition (\citealp{deng2013new}), computer vision (\citealp{girshick2015fast}) and drug discovery (\citealp{Ramsundar2015MassivelyMN}). But MTL is less commonly used and has seen limited success in BioNER so far. \citeauthor{korhonen2017neural} explored MTL with a CNN model for BioNER. However, \citeauthor{korhonen2017neural} only considers word-level features as input, ignoring character-level lexical information which are often crucial for modeling biomedical entities (e.g. -ase could be an important subword feature for gene/protein entity recognition). As a result, their best performing multi-task CNN model does not outperform state-of-the-art systems that use on handcrafted features (\citealp{korhonen2017neural}).
	
	In this paper, we propose a new multi-task learning framework using char-level neural models for BioNER. The proposed framework, despite being simple and not requiring any feature engineering, achieves excellent benchmark performance. Our multi-task model is built upon a single-task neural network model (\citealp{2017arXiv170904109L}). In particular, we consider a BiLSTM-CRF model with an additional context-dependent BiLSTM layer for modeling character sequences. A prominent advantage of our multi-task model is that inputs from different datasets can efficiently share both character- and word-level representations, by reusing parameters in the corresponding BiLSTM units.
	We compare the proposed multi-task model with state-of-the-art BioNER systems and baseline neural network models on 15 benchmark BioNER datasets and observe substantially better performance. We further show through detailed experimental analysis on 5 datasets that the proposed approach adds marginal computational overhead and outperforms strong baseline neural models that do not consider multi-task learning, suggesting that multi-task learning plays an important role in its success. Altogether, this work introduces a new text-mining approach that can help scientists exploit knowledge buried in biomedical literature in a systematic and unbiased way.
	

	\begin{methods}
		\vspace{-1em}
		\section{Background}  \label{models}
		
		\subsection{NER problem definition}
		Let $\Phi$ denote the set of labels indicating whether a word is part of a specific entity type or not. Given a sequence of words $\boldsymbol{w} = \{w_1, w_2, ..., w_n\}$, the output is a sequence of labels $\boldsymbol{y} = \{y_1, y_2, ..., y_n\}$, $y_i\in\Phi$. \xuan{For example, given a sentence ``.... including the RING1 ...", the output should be ``... O O S-GENE ..." in which ``O" indicates a non-entity type and ``S-GENE" indicates a single-token GENE type.}
		
		\subsection{Long Short-Term Memory (LSTM)} 
		Long short-term memory neural network is a specific type of recurrent neural network that models dependencies between elements in a sequence through recurrent connections (Fig. \ref{lstm}). The input to an LSTM network is a sequence of vectors $\boldsymbol{X} = \{\boldsymbol{x}_1, \boldsymbol{x}_2, ..., \boldsymbol{x}_T\}$, where vector $\boldsymbol{x}_i$ is a representation vector of a word in the input sentence. The output is a sequence of vectors $\boldsymbol{H} = \{\boldsymbol{h}_1, \boldsymbol{h}_2, ..., \boldsymbol{h}_T\}$, where $\boldsymbol{h}_i$ is a hidden state vector. At step $t$ of the recurrent calculation, the network takes $\boldsymbol{x}_t, \boldsymbol{c}_{t-1}, \boldsymbol{h}_{t-1}$ as inputs and produces $\boldsymbol{c}_t, \boldsymbol{h}_t$ via the following intermediate calculations:
		\begin{equation*}
		\boldsymbol{i}_t = \sigma(\boldsymbol{W}^i\boldsymbol{x}_t + \boldsymbol{U}^i\boldsymbol{h}_{t-1} + \boldsymbol{b}^i)
		\end{equation*}
		\begin{equation*}
		\boldsymbol{f}_t = \sigma(\boldsymbol{W}^f\boldsymbol{x}_t + \boldsymbol{U}^f\boldsymbol{h}_{t-1} + \boldsymbol{b}^f)
		\end{equation*}
		\begin{equation*}
		\boldsymbol{o}_t = \sigma(\boldsymbol{W}^o\boldsymbol{x}_t + \boldsymbol{U}^o\boldsymbol{h}_{t-1} + \boldsymbol{b}^o)
		\end{equation*}
		\begin{equation*}
		\boldsymbol{g}_t = tanh(\boldsymbol{W}^g\boldsymbol{x}_t + \boldsymbol{U}^g\boldsymbol{h}_{t-1} + \boldsymbol{b}^g)
		\end{equation*}
		\begin{equation*}
		\boldsymbol{c}_t =\boldsymbol{f}_t\odot\boldsymbol{c}_{t-1} + \boldsymbol{i}_t\odot\boldsymbol{g}_t
		\end{equation*}
		\begin{equation*}
		\boldsymbol{h}_t =\boldsymbol{o}_t\odot tanh(\boldsymbol{c}_t),
		\end{equation*}
		where $\sigma(\cdot)$ and $tanh(\cdot)$ denote element-wise sigmoid and hyperbolic tangent functions, respectively, and $\odot$ denotes element-wise multiplication. The $\boldsymbol{i}_t$, $\boldsymbol{f}_t$ and $\boldsymbol{o}_t$ are referred to as input, forget, and output gates, respectively. The $\boldsymbol{g}_t$ and $\boldsymbol{c}_t$ are intermediate calculation steps. At $t = 1$, $\boldsymbol{h}_0$ and $\boldsymbol{c}_0$ are initialized to zero vectors. The trainable parameters are $\boldsymbol{W}^j$,$\boldsymbol{U}^j$ and $\boldsymbol{b}^j$ for $j\in \{i, f, o, g\}$.
		
		The LSTM architecture described above can only process the input in one direction. The bi-directional long short-term memory (BiLSTM) model improves the LSTM by feeding the input to the LSTM network twice, once in the original direction and once in the reversed direction. Outputs from both directions are concatenated to represent the final output. This design allows for detection of dependencies from both previous and subsequent words in a sequence.
		
		\begin{figure}[!tpb]
			\centerline{\includegraphics[trim={0.5cm 6.5cm 1.5cm 0.5cm},clip, width=0.5\textwidth]{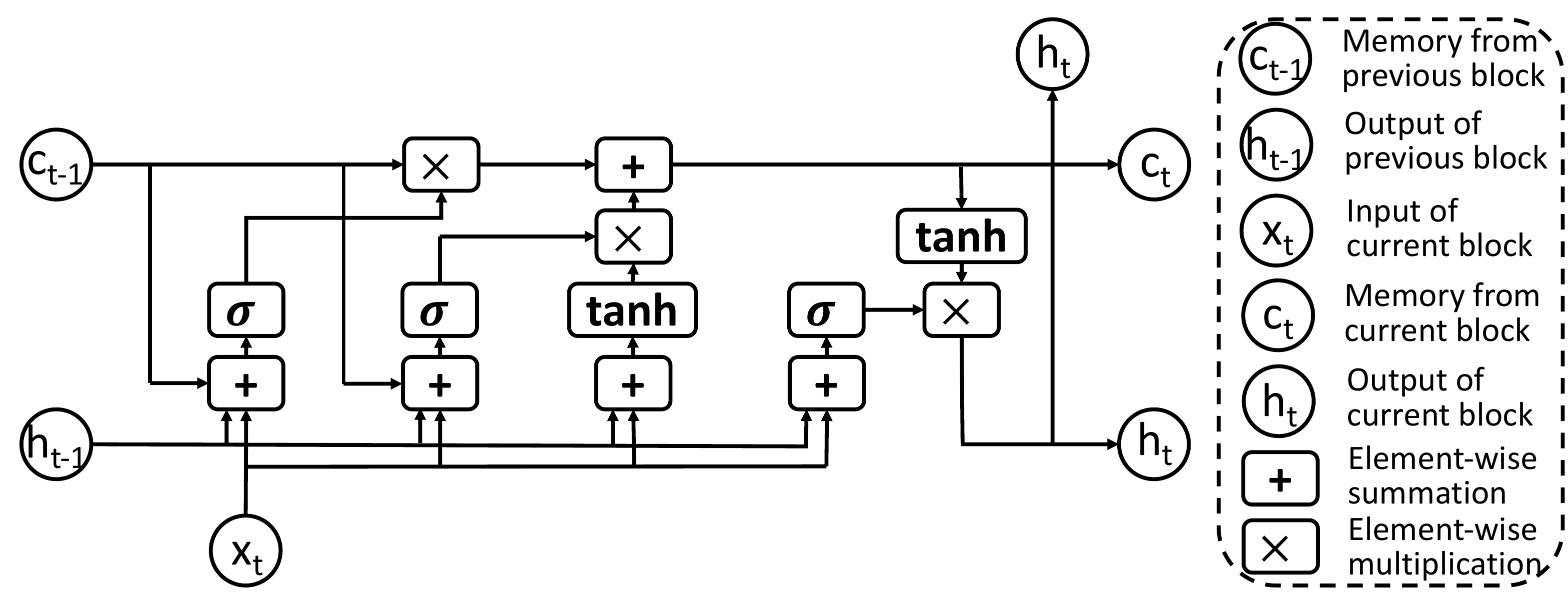}}
			\vspace{-1em}
			\caption{Architecture of long short-term memory neural network.}\label{lstm}
			\vspace{-2em}
		\end{figure}
		
		\begin{figure*}[!tpb]
			\centerline{\includegraphics[trim={1cm 1cm 2cm 6cm},clip, width=\textwidth]{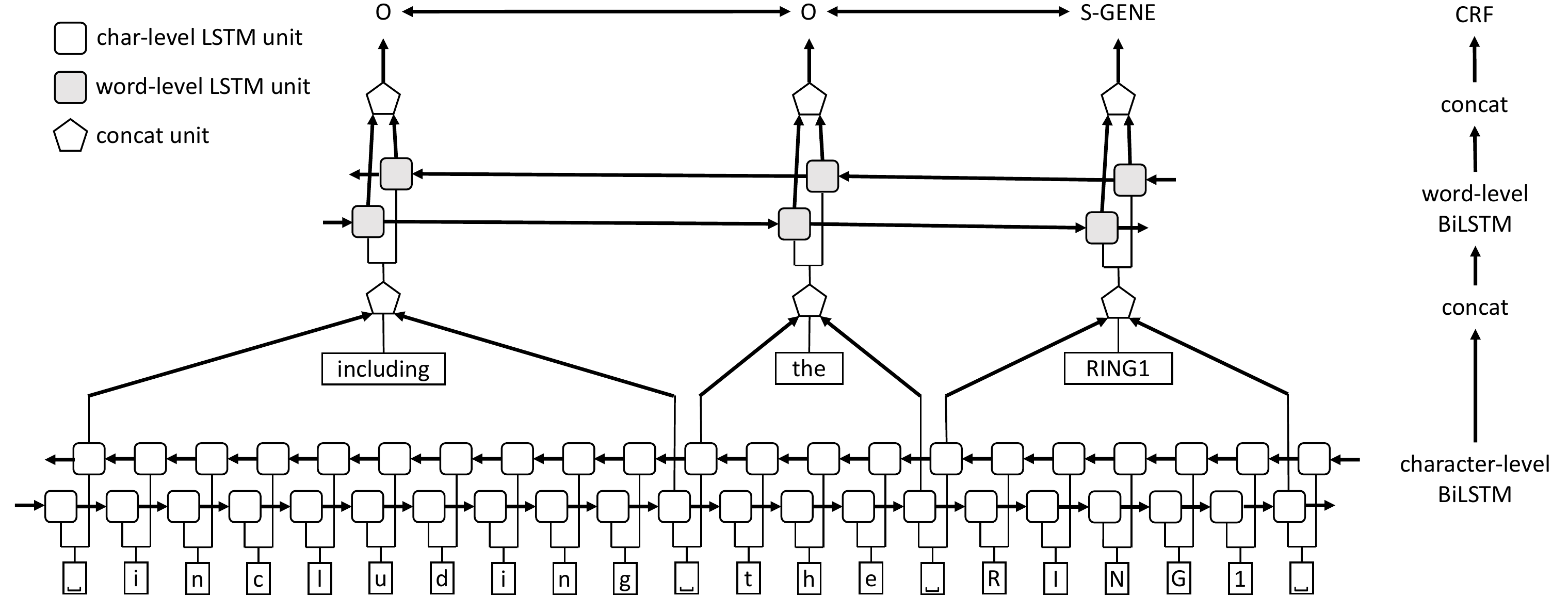}}
			\vspace{-1em}
			\caption{\xuan{Architecture of a single-task neural network. The input is a sentence from biomedical literature. Rectangles denote character and word embeddings; empty round rectangles denote the first character-level BiLSTM; shaded round rectangles denote the second word-level BiLSTM; pentagons denote the concatenation units. The tags on the top, e.g., 'O', 'S-GENE', are the output of the final CRF layer, which are the entity labels we get for each word in the sentence.}}\label{stm}
		\end{figure*}
		
		\subsection{Bi-directional Long Short-Term Memory-Conditional Random Field (BiLSTM-CRF)} \label{bilstm-crf}
		A naive way of applying the BiLSTM network to sequence labeling is to use the output hidden state vectors to make independent tagging decisions. However, in many sequence labeling tasks such as BioNER, it is useful to also model the dependencies across output tags. The BiLSTM-CRF network adds a conditional random field (CRF) layer on top of a BiLSTM network. This BiLSTM-CRF network takes the input sequence $\boldsymbol{X} = \{\boldsymbol{x}_1, \boldsymbol{x}_2, ..., \boldsymbol{x}_n\}$ to predict an output label sequence $\boldsymbol{y} = \{y_1, y_2, ..., y_n\}$. A score is defined as:
		\begin{equation*}
		s(\boldsymbol{X}, \boldsymbol{y}) = \sum_{i=0}^{n}\boldsymbol{A}_{y_i, y_{i+1}} + \sum_{i=1}^{n}\boldsymbol{P}_{i, y_i},
		\end{equation*}
		where $\boldsymbol{P}$ is an $n\times k$ matrix of the output from the BiLSTM layer, $n$ is the sequence length, $k$ is the number of distinct labels, $\boldsymbol{A}$ is a $(k+2)\times (k+2)$ transition matrix and $\boldsymbol{A}_{i, j}$ represents the transition probability from the $i$-th label to the $j$-th label. Note that two additional labels \textit{<start>} and \textit{<end>} are used to represent the start and end of a sentence, respectively. We further define $\boldsymbol{Y}_{\boldsymbol{X}}$ as all possible sequence labels given the input sequence $\boldsymbol{X}$. The training process maximizes the log-probability of the label sequence $\boldsymbol{y}$ given the input sequence $\boldsymbol{X}$:
		\begin{equation}\label{crf}
		log(p(\boldsymbol{y}|\boldsymbol{X})) = log\frac{e^{s(\boldsymbol{X}, \boldsymbol{y})}}{\sum_{\boldsymbol{y}'\in \boldsymbol{Y}_{\boldsymbol{X}}}e^{s(\boldsymbol{X}, \boldsymbol{y'})}}.
		\end{equation}
		
		A three-layer BiLSTM-CRF architecture is employed by \citeauthor{lample2016neural} and \citeauthor{habibi2017deep} to jointly model the word and the character sequences in the input sentence. In this architecture, the first BiLSTM layer takes character embedding sequence of each word as input, and produces a character-level representation vector for this word as output. This character-level vector is then concatenated with a word embedding vector, and fed into a second BiLSTM layer. Lastly, a CRF layer takes the output vectors from the second BiLSTM layer, and outputs the best tag sequence by maximizing the log-probability in Equation~\ref{crf}.
		
		In practice, the character embedding vectors are randomly initialized and co-trained during the model training process. The word embedding vectors are retrieved directly from a pre-trained word embedding lookup table. The classical Viterbi algorithm is used to infer the final labels for the CRF model. The three-layer BiLSTM-CRF model is a differentiable neural network architecture that can be trained by backpropagation.\\
		
		\begin{figure*}[!tpb]
			\centering
			\subfigure[MTM-C]{%
				\label{mtm1}%
				\includegraphics[trim={3cm 5cm 19cm 3cm},clip, scale=0.35]{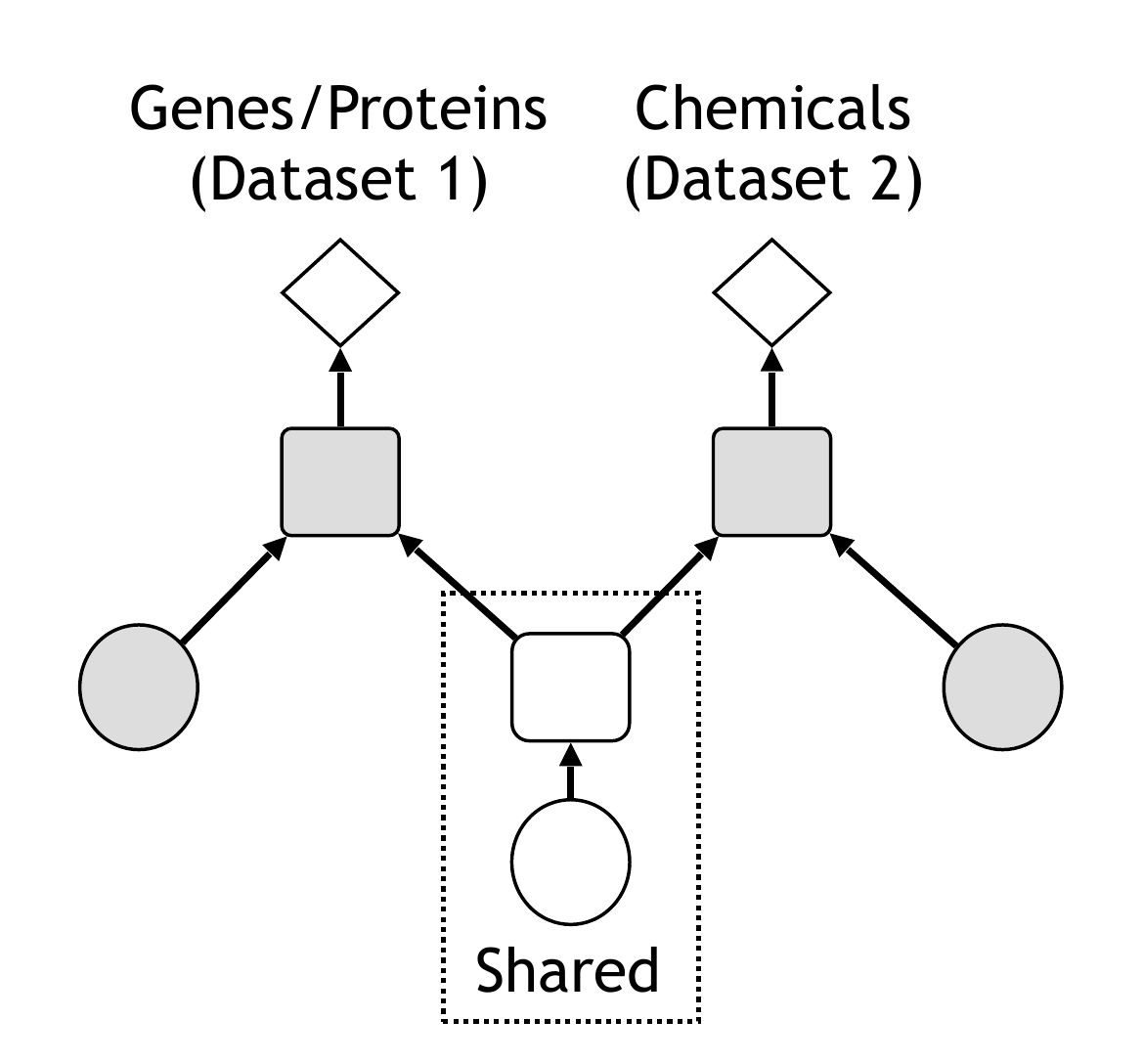}}%
			\qquad
			\subfigure[MTM-W]{%
				\label{mtm2}%
				\includegraphics[trim={3cm 5cm 21cm 3cm},clip, scale=0.35]{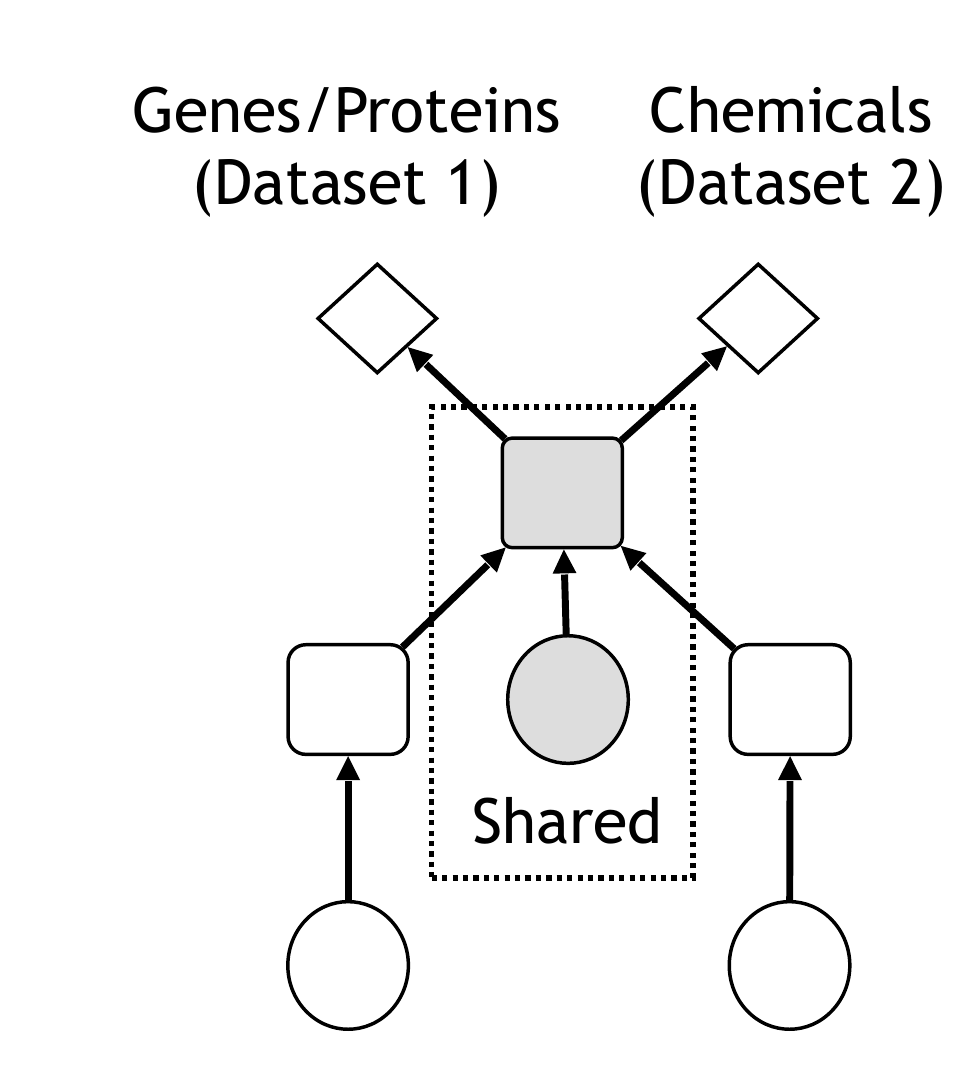}}%
			\qquad
			\subfigure[MTM-CW]{%
				\label{mtm3}%
				\includegraphics[trim={3cm 5cm 21cm 3cm},clip, scale=0.35]{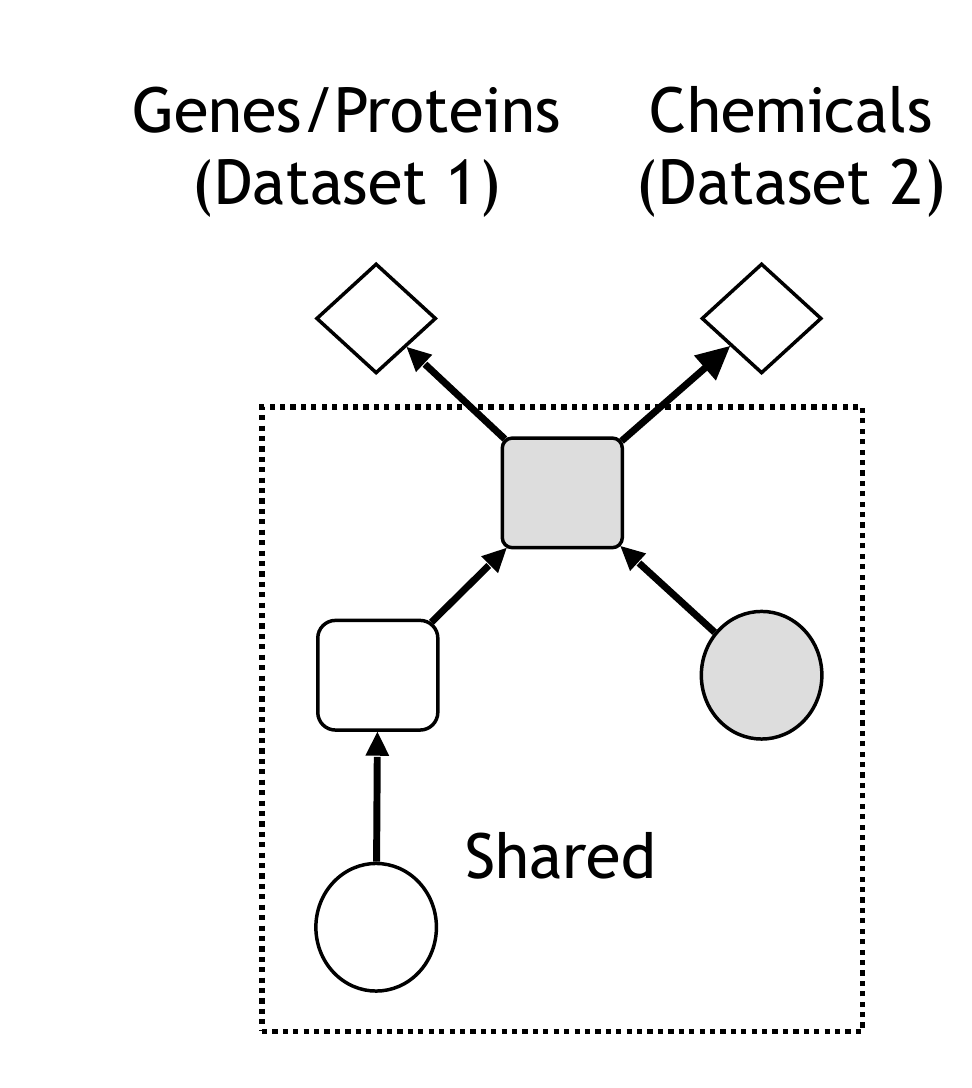}}%
			\qquad
			\subfigure{%
				\label{mtm_legend}%
				\includegraphics[trim={13cm 8cm 13cm 3cm},clip, scale=0.45]{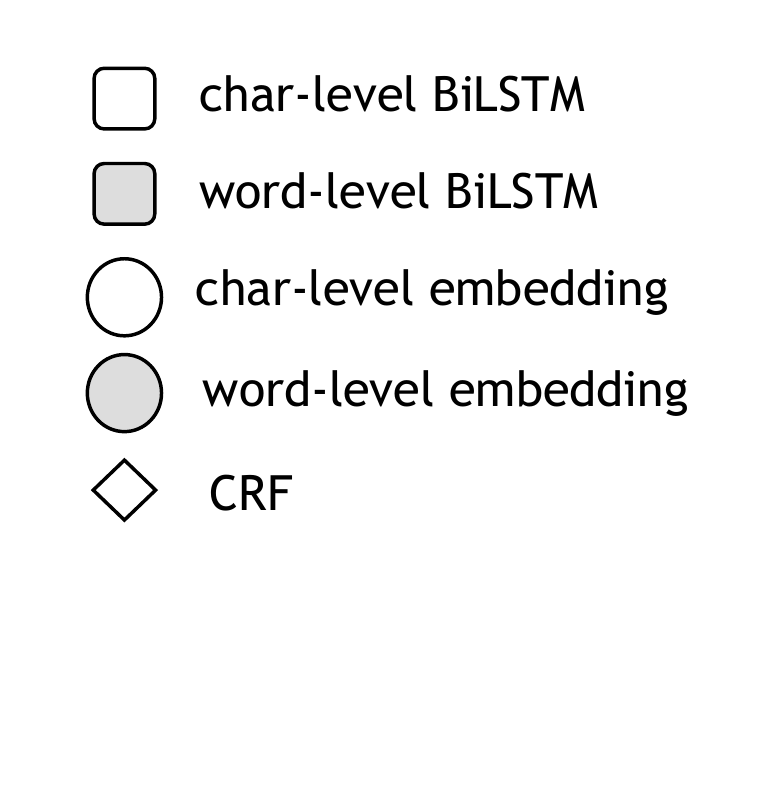}}%
			\vspace{-1em}
			\caption{\xuan{Three multi-task learning neural network models. The empty circles denote the character embeddings. The empty round rectangles denote the character-level BiLSTM. The shaded circles denote the word-level embeddings. The shaded round rectangles denote the word-level BiLSTM. The squares denote the CRF layer. (a) MTM-C: multi-task learning neural network with a shared character layer and a task-specific word layer, (b): MTM-W: multi-task learning neural network with a task-specific character layer and a shared word layer, (c) MTM-CW: multi-task learning neural network with shared character and word layers.}}
			\vspace{-2em}
			\label{mtm1-3}
		\end{figure*}
		
		\vspace{-2em}
		\section{Deep multi-task learning for BioNER}
		
		\subsection{Single-task model (STM)} \label{sec:single-task}
		The vanilla BiLSTM-CRF model can learn high-quality representations for words that appeared in the training dataset. However, it often fails to generalize to out-of-vocabulary (OOV) words (i.e., words that did not appear in the training dataset) because they don't have a pre-trained word embedding. These OOV words are common in biomedical text (67.21\% OOV words of the datasets in Table \ref{dataset}). Therefore, for the baseline single-task BioNER model, we use a neural network architecture that better handles OOV words. As shown in Fig. \ref{stm}, our single-task model consists of three layers. In the first layer, a BiLSTM network is used to model the character sequence of the input sentence. We use character embedding vectors as input to the network. Hidden state vectors at the word boundaries of this character-level BiLSTM are then selected and concatenated with word embedding vectors to form word representations. Next, these word representation vectors are fed into a word-level BiLSTM layer (i.e., the upper BiLSTM layer in Fig. \ref{stm}). Lastly, output of this word-level BiLSTM is fed into the a CRF layer for label prediction. Compared to the vanilla BiLSTM-CRF model, a major advantage of this model is that it can infer the meaning of an out-of-vocabulary word from its character sequence and other characters around it. For example, the model is now able to infer that ``RING2'' likely represents a gene symbol, even though then network may have only seen the word ``RING1'' during training.
		
		\subsection{Multi-task models (MTMs)} \label{sec:multi-task}
		An important characteristic of the BioNER task is the limited availability of supervised training data. We propose a multi-task learning approach to address this problem by training different BioNER models on datasets with different entity types while sharing parameters across these models. We hypothesize that the proposed approach can make more efficient use of the data and encourage the models to learn representations of words and characters (which are shared between multiple corpora) in a more effective and generalized way.
		
		We give a formal definition of the multi-task setting as the following. Given $m$ datasets, for $i\in\{1,...,m\}$, each dataset $D_i$ consists of $n_i$ training samples, i.e., $D_i = \{\boldsymbol{w}_j^i, y_j^i\}_{j=1}^{n_i}$. We denote the training matrix for each dataset as $\boldsymbol{X}^i = \{\boldsymbol{x}_1^i, ..., \boldsymbol{x}_{n_i}^i\}$ ($\boldsymbol{X}^i$ is the feature representation of the input word sequence $\boldsymbol{w}_j^i$) and the labels for each dataset as $\boldsymbol{y}^i = \{y_1^i, ..., y_{n_i}^i\}$. \xuan{The model parameters include the word-level BiLSTM parameters ($\theta_i^w$), the character-level BiLSTM parameters ($\theta_i^c$) and the output CRF parameters ($\theta_i^{o}$).} A multi-task model therefore consists of $m$ different models, each trained on a separate dataset, while sharing part of the model parameters across datasets. The loss function $L$ of the multi-task model is:
		\begin{equation*}
		L = \sum_{i=1}^{m}\lambda_iL_i = \sum_{i=1}^{m}\lambda_i \log(P_{\theta_i^w, \theta_i^c,\theta_i^{o}}(\boldsymbol{y}^i|\boldsymbol{X}^i)).
		\end{equation*}
		The log-likelihood term is shown in Equation~\ref{crf} and $\lambda_i$ is a positive hyper-parameter that controls the contribution of each dataset. 
		We observed that our multi-task model is able to achieve very competitive performance with $\lambda_i = 1$ on all datasets that we evaluated on and therefore use this value in our experiments. However, we believe that the performance can be improved with further tuned $\lambda_i$ values.
		
		We propose three different multi-task models, as illustrated in Fig. \ref{mtm1-3}. These three models differ in which part of the model parameters ($\theta_i^w, \theta_i^c,\theta_i^{o}$) are shared across multiple datasets:
		
		\paragraph{\textbf{MTM-C}}
		In this model, $\theta^c_i = \theta^c$ are shared among different tasks. All datasets are iteratively used to train the model. When a dataset is used, the parameters updated during the training are $\theta^c$ and $\theta^w_i$. The detailed architecture of this multi-task model is shown in Fig. \ref{mtm1}.
		
		\paragraph{\textbf{MTM-W}}
		In this model, $\theta^w_i = \theta^w$ are shared among different tasks. When a dataset is used, the parameters updated during the training are $\theta^w$ and $\theta^c_i$.  The detailed architecture of this multi-task model is shown in Fig. \ref{mtm2}.
		
		\paragraph{\textbf{MTM-CW}}
		In this model, $\theta^c_i = \theta^c$ and $\theta^w_i = \theta^w$ are shared among different tasks. Each dataset has its specific $\theta^{o}_i$ for label prediction. MTM-CW shared the most information across tasks compared with the other two multi-task models. It enables sharing both character- and word-level information between different biomedical entities, while the other two models only enable sharing part of the information. The detailed architecture of this multi-task model is shown in Fig. \ref{mtm3}. 
		
		%
		%
		
		\vspace{-1em}
		\section{Experimental setup}
		
		\subsection{Datasets} \label{datasets}
		We test our method on the same 15 datasets used by \citeauthor{korhonen2017neural}, and find our model achieves substantially better performance on 14 of them compared with baseline neural network models. Due to space limit, here we report detailed results of the multi-task model on 5 main datasets (Table \ref{dataset}), which altogether cover major biomedical entity types (e.g., genes, proteins, chemicals, diseases). We also include full results on all the 15 datasets in \emph{Supplementary Material: Performance comparison on 15 datasets}. The performance of the multi-task model is slightly different when trained on 5 datasets compared with trained on 15 datasets (shown in \emph{Supplementary Material: Performance comparison on 15 datasets}), as the MTL model has access to more data. In our experiments, we follow the experiment setup of \citeauthor{korhonen2017neural} and divide each dataset into training, development and test sets. We use training and development sets to train the final model. All datasets are publicly available.\footnote{All datasets can be downloaded from: \\ \href{https://github.com/cambridgeltl/MTL-Bioinformatics-2016}{https://github.com/cambridgeltl/MTL-Bioinformatics-2016.}} As part of preprocessing, word labels are encoded using an IOBES scheme. In this scheme, for example, a word describing a gene entity is tagged with ``B-Gene'' if it is at the beginning of the entity, ``I-Gene'' if it is in the middle of the entity, and ``E-Gene'' if it is at the end of the entity. Single-word gene entities are tagged with ``S-Gene''. All other words not describing entities of interest are tagged as `O'. Next, we briefly describe the 5 main datasets and their corresponding state-of-the-art BioNER systems.
		
		\begin{table}[t]
			\setlength{\tabcolsep}{2pt}
			\centering
			\caption{Biomedical NER datasets used in the experiments.}
			\label{dataset}
			\begin{tabular}{lll}\hline
				Dataset & Size & Entity types and counts \\ \hline
				BC2GM & 20,000 sentences & Gene/Protein (24,583)\\
				BC4CHEMD & 10,000 abstracts & Chemical (84,310)\\
				BC5CDR & 1,500 articles & Chemical (15,935), Disease (12,852)\\
				NCBI-Disease & 793 abstracts & Disease (6,881)\\
				\multirow{3}{*}{JNLPBA} & \multirow{3}{*}{2,404 abstracts} & Gene/Protein (35,336),  \\
				& & Cell Line (4,330), DNA (10,589), \\ 
				& & Cell Type (8,649), RNA (1,069)\\\hline
			\end{tabular}
			\vspace{-2em}
		\end{table}
		
		\begin{table*}[t]
			\centering
			\caption{Performance and average training time of the baseline neural network models and the proposed MTM-CW model. \textbf{Bold}: best scores, *: significantly worse than the MTM-CW model ($p\leq 0.05$), **: significantly worse than the MTM-CW model ($p\leq 0.01$). The details of dataset benchmark systems and evaluation methods are described in Section \ref{datasets} and \ref{evaluation}, respectively. }
			\label{performance}
			\begin{tabular}{llp{1.9cm}p{1.9cm}p{1.9cm}p{1.9cm}p{1.9cm}p{1.9cm}}
				\hline
				\textbf{}                                                                    & \textbf{} & \begin{tabular}[c]{@{}l@{}}Dataset \\ Benchmark\end{tabular} & Crichton \textit{et al.}  & \begin{tabular}[c]{@{}l@{}}Lample \textit{et al.}\\ Habibi \textit{et al.} \end{tabular} & Ma and Hovy      & STM            & MTM-CW             \\ \hline
				\multirow{3}{*}{\begin{tabular}[c]{@{}l@{}}BC2GM\\ (Exact)\end{tabular}}       &  Precision    & -                                                            & -               & 81.57$\pm$0.26$^{*}$                                                                 & 79.09$\pm$0.63$^{**}$            & 81.11$\pm$0.33$^{*}$          & \textbf{82.10$\pm$0.04} \\
				& Recall      & -                                                            & -               & \textbf{79.48$\pm$0.27}                                                                 & 77.87$\pm$0.53$^{**}$            & 78.91$\pm$0.40$^{**}$          & 79.42$\pm$0.01   \\
				& F1          & -                                                            & 73.17$^{**}$    & 80.51$\pm$0.09                                                          & 78.48$\pm$0.31$^{**}$     & 80.00$\pm$0.15$^{*}$    & \textbf{80.74$\pm$0.04} \\ \hline
				\multirow{3}{*}{\begin{tabular}[c]{@{}l@{}}BC2GM\\ (Alternative)\end{tabular}} & Precision    & 88.48                                                        & -               & 87.27$\pm$0.41$^{**}$                                                                 &   83.50$\pm$0.37$^{**}$               & 88.21$\pm$0.28$^{*}$          & \textbf{89.45$\pm$0.32}   \\
				& Recall      & 85.97$^{**}$                                                        & -               & 87.84$\pm$0.19                                                                 &    87.13$\pm$0.17$^{*}$              & 87.43$\pm$0.18$^{*}$          & \textbf{88.67$\pm$0.37}   \\
				& F1          & 87.21$^{**}$                                                        & 84.41$^{**}$           & 87.55$\pm$0.10$^{*}$                                                                 &   85.27$\pm$0.11$^{**}$               & 87.82$\pm$0.30$^{*}$          & \textbf{89.06$\pm$0.32}   \\ \hline
				\multirow{3}{*}{BC4CHEMD}                                                      & Precision   & 88.73$^{**}$                                                        & -               & 89.68$\pm$0.22$^{*}$                                                                 & 90.83$\pm$0.53 & 90.53$\pm$0.72$^{*}$          & \textbf{91.30$\pm$0.08}           \\
				& Recall      & 87.41                                                        & -               & 85.87$\pm$0.16$^{*}$                                                                & 83.19$\pm$0.20$^{**}$            & 87.04$\pm$0.50          & \textbf{87.53$\pm$0.11} \\
				& F1          & 88.06$^{*}$                                                        & 83.02$^{**}$    & 87.74$\pm$0.05$^{**}$                                                          & 86.84$\pm$0.07$^{**}$     & 88.75$\pm$0.20   & \textbf{89.37$\pm$0.07} \\ \hline
				\multirow{3}{*}{BC5CDR}                                                        & Precision   & \textbf{89.21}                                               & -               & 87.60$\pm$0.08$^{**}$                                                                 & 89.16$\pm$0.03            & 88.84$\pm$0.08         & 89.10$\pm$0.11            \\
				& Recall      & 84.45$^{**}$                                                        & -               & 86.25$\pm$0.07$^{**}$                                                                 & 84.28$\pm$0.02$^{**}$            & 85.16$\pm$0.05$^{**}$          & \textbf{88.47$\pm$0.04} \\
				& F1          & 86.76$^{**}$                                                        & 83.90$^{**}$    & 86.92$\pm$0.06$^{**}$                                                                 & 86.65$\pm$0.06$^{**}$     & 86.96$\pm$0.00$^{**}$   & \textbf{88.78$\pm$0.12} \\ \hline
				\multirow{3}{*}{NCBI-Disease}                                                          & Precision   & 85.10                                                        & -               &86.11$\pm$0.33                                                      &  \textbf{86.89$\pm$0.34}            & 84.95$\pm$0.41          & 85.86$\pm$0.90           \\
				& Recall      & 80.80$^{**}$                                                        & -               & 85.49$\pm$0.93                                                                 & 78.75$\pm$0.16$^{**}$            & 82.92$\pm$0.31$^{*}$          & \textbf{86.42$\pm$0.44} \\
				& F1          & 82.90$^{**}$                                                        & 80.37$^{**}$    & 85.80$\pm$0.16                                                          & 82.62$\pm$0.29$^{**}$      & 83.92$\pm$0.18$^{*}$          & \textbf{86.14$\pm$0.31} \\ \hline
				\multirow{3}{*}{JNLPBA}                                                        & Precision   & 69.42$^{**}$                                                        & -               & \textbf{71.35$\pm$0.05}                                                                 & 70.28$\pm$0.03$^{*}$            & 69.60$\pm$0.07$^{**}$          & 70.91$\pm$0.02 \\
				& Recall      & 75.99                                                        & -               & 75.74$\pm$0.07                                                                 & 75.26$\pm$0.41            & 74.95$\pm$0.24$^{*}$ & \textbf{76.34$\pm$0.23}            \\
				& F1          & 72.55$^{**}$                                                        & 70.09$^{**}$    & 73.48$\pm$0.03                                                          & 72.68$\pm$0.21$^{*}$      & 72.17$\pm$0.13$^{**}$          & \textbf{73.52$\pm$0.03} \\ \hline
				\textbf{Training time (s/sent.)} & & - & - & 1.59 & 0.95 & 0.71 & 0.75 \\\hline
			\end{tabular}
			\vspace{-2em}
		\end{table*}
		
		\paragraph{\textbf{BC2GM}}
		The state-of-the-art system reported for the BioCreative II gene mention recognition task adopts semi-supervised learning method with alternating structure optimization (\citealp{ando2007biocreative}).
		
		\paragraph{\textbf{BC4CHEMD}}
		The state-of-the-art system reported for the BioCreative IV chemical entity mention recognition task is the \emph{CHEMDNER} system (\citealp{lu2015chemdner}), which is based on mixed conditional random fields with Brown clustering of words.
		
		\paragraph{\textbf{BC5CDR}}
		The state-of-the-art system reported for the most recent BioCreative V chemical and disease mention recognition task is the \emph{TaggerOne} system (\citealp{leaman2016taggerone}), which uses a semi-Markov model for joint entity recognition and normalization.
		
		\paragraph{\textbf{NCBI-Disease}}
		The NCBI disease dataset was initially introduced for disease name recognition and normalization. It has been widely used for a lot of applications. The state-of-the-art system on this dataset is also the \emph{TaggerOne} system (\citealp{leaman2016taggerone}).
		
		\paragraph{\textbf{JNLPBA}}
		The state-of-the-art system (\citealp{guodong2004exploring}) for the 2004 JNLPBA shared task on biomedical entity (gene/protein, DNA, RNA, cell line, cell type) recognition uses a hidden markov model (HMM). Although this task and the model is a bit old compared with the others, it still remains a competitive benchmark method for comparison.
		
		\subsection{Evaluation metrics}\label{evaluation}
		We report the performance of all the compared methods on the test set. We deem each predicted entity as correct only if both the entity boundary and entity types are the same as the ground-truth annotation (i.e., \emph{exact match}). Then we calculate the precision, recall and F1 scores on all datasets and macro-averaged F1 scores on all entity types. For error analysis, we compare the ratios of false positive (FP) and false negative (FN) labels in the single-task and the multi-task models and include the results in \emph{Supplementary Material: Error analysis}.
		
		The test set of the BC2GM dataset is constructed slightly differently compared to the test sets of other datasets. BC2GM additionally provides a list of alternative answers for each entity in the test set. A predicted entity is deemed correct as long as it matches the ground truth or one of the alternative answers. We refer to this measurement as \emph{alternative match} and report scores under both \emph{exact match} and \emph{alternative match} for the BC2GM dataset. 
		
		\subsection{Pre-trained word embeddings}
		We initialize the word embedding matrix with pre-trained word vectors from \citealp{moen2013distributional} in all experiments.\footnote{The pre-trained word vectors can be download from: \\ \href{http://bio.nlplab.org/}{http://bio.nlplab.org/}.} These word embeddings are trained using a skip-gram model, as described in \citealp{mikolov2013efficient}. These word vectors are trained on three different datasets: (1) abstracts from the PubMed database, (2) abstracts from the PubMed database together with full-text articles from the PubMed Central (PMC), and (3) the entire Pubmed database of abstracts and full-text articles together with the Wikipedia corpus. We found the third set of word vectors lead to best results on development set and therefore used it for the model development. We provide a full comparison of different word embeddings in \emph{Supplementary Material: Performance of Word Embeddings}. In all experiments, we replace rare words (i.e., words with a frequency of less than 5) with a special \emph{<UNK>} token, whose embedding is randomly initialized and fine-tuned during model training.
		
		\subsection{Training details}
		All the neural network models are trained on one GeForce GTX 1080 GPU. To train our neural models, we use a learning rate of 0.01 with a decay rate of 0.05 applied to every epoch of training. The dimensions of word and character embedding vectors are set to be 200 and 30, respectively (\citealp{2017arXiv170904109L}). We adopted 200 (best performance among 100, 200 and 300) for both character- and word-level BiLSTM layers. Note that \citeauthor{2017arXiv170904109L} considers advanced strategies, such as highway structures, to further improve performance. We did not observe any significant performance boost with these advanced strategies, thus do not adopt these strategies in this work. The performance of the model variations with these advanced strategies can be found in \emph{Supplementary Material: Performance of Model Variations}.
		To train the baseline neural network models, we use the default parameter settings as used in their paper (\citealp{lample2016neural, habibi2017deep, ma2016end}) because we found the default parameters also lead to almost optimal performance on the development set.
		
		\vspace{-1em}
		\section{Results}
		\subsection{Performance comparison on benchmark datasets}\label{sec:datasets}
		We compare the proposed single-task (Section~\ref{sec:single-task}) and multi-task models (Section~\ref{sec:multi-task}) with state-of-the-art BioNER systems (reported for each dataset) and three neural network models from \citeauthor{korhonen2017neural}, \citeauthor{lample2016neural, habibi2017deep}, and \citeauthor{ma2016end}. The evaluation metrics include precision, recall and F1 score (\citealp{tsai2006various}) (Table \ref{performance}). We denote results of the best system priorly reported for each dataset as ``Dataset Benchmark". For method proposed by \citeauthor{korhonen2017neural}, we quote their experiment results directly. For other neural network models, we repeart each experiment three times with the mean and standard deviation reported (Table \ref{performance}). To directly compare with the results in \citeauthor{korhonen2017neural}, we measure statistical significance with the same t-test as used in their paper.
		
		We observe that the MTM-CW model achieves significantly higher F1 scores than state-of-the-art benchmark systems (column Dataset Benchmark in Table \ref{performance}) on all of the five datasets. Following established practice in the literature, we use exact match to compare benchmark performance on all the datasets except for the BC2GM, where we report benchmark performance based on alternative match. Furthermore, MTM-CW generally achieves significantly higher F1 scores than other neural network models. 
		These results show that the proposed multi-task learning neural network significantly outperforms state-of-the-art systems and other neural networks. In particular, the MTM-CW model consistently achieves a better performance than the single task model, demonstrating that multi-task learning is able to successfully leverage information across different datasets and mutually enhance performance on each single task. We further investigate the performance of three multi-task models (MTM-C, MTM-W, and MTM-CW, Table \ref{multi}). Results show that the best performing multi-task model is MTM-CW, indicating the importance of morphological information captured by character-level BiLSTM as well as lexical and contextual information captured by word-level BiLSTM.
		
		\begin{table}[]
			\centering
			\caption{F1 scores of three multi-task models proposed in this paper. \textbf{Bold}: best scores, *: significantly worse than the MTM-CW model ($p\leq 0.05$), **: significantly worse than the MTM-CW model ($p\leq 0.01$).}
			\label{multi}
			\begin{tabular}{p{2cm}p{1.7cm}p{1.7cm}p{1.7cm}}
				\hline
				Dataset  & MTM-C  & MTM-W  & MTM-CW           \\ \hline
				BC2GM    & 77.80$\pm$0.30$^{**}$ & 79.42$\pm$0.08$^{**}$ & \textbf{80.74$\pm$0.04} \\
				BC4CHEMD & 88.16$\pm$0.07$^{*}$ & 88.49$\pm$0.03$^{*}$ & \textbf{89.37$\pm$0.07} \\
				BC5CDR   & 86.05$\pm$0.26$^{**}$ & 88.26$\pm$0.05$^{*}$ & \textbf{88.78$\pm$0.12} \\
				NCBI-Disease     & 82.94$\pm$0.31$^{**}$ & 84.81$\pm$0.14 & \textbf{86.14$\pm$0.31} \\
				JNLPBA   & 71.79$\pm$0.41$^{**}$ & 73.21$\pm$0.16 & \textbf{73.52$\pm$0.03} \\ \hline
			\end{tabular}
			\vspace{-0.5em}
		\end{table}

		\subsection{Performance on major biomedical entity types}
		We also conduct more fine-grained comparison of all models on four major biomedical entity types: genes/proteins, chemicals, diseases and cell lines since they are the most often annotated entity types (Fig. \ref{bmentity}). Each entity type comes from multiple datasets: genes/proteins from BC2GM and JNLPBA, chemicals from BC4CHEMD and BC5CDR, diseases from BC5CDR and NCBI-Disease, and cell lines from JNLPBA.
		
		The MTM-CW model performs consistently better than the neural network model (\citealp{habibi2017deep}) on all four entity types. It also outperforms the state-of-the-art systems (Benchmark in Fig. \ref{bmentity}) on three entity types except for cell lines. These results further confirm that the multi-task neural network model achieves a significantly better performance compared with state-of-art systems and other neural network models for BioNER.
		
		\begin{figure}[!tpb]
			\centerline{\includegraphics[width=0.5\textwidth]{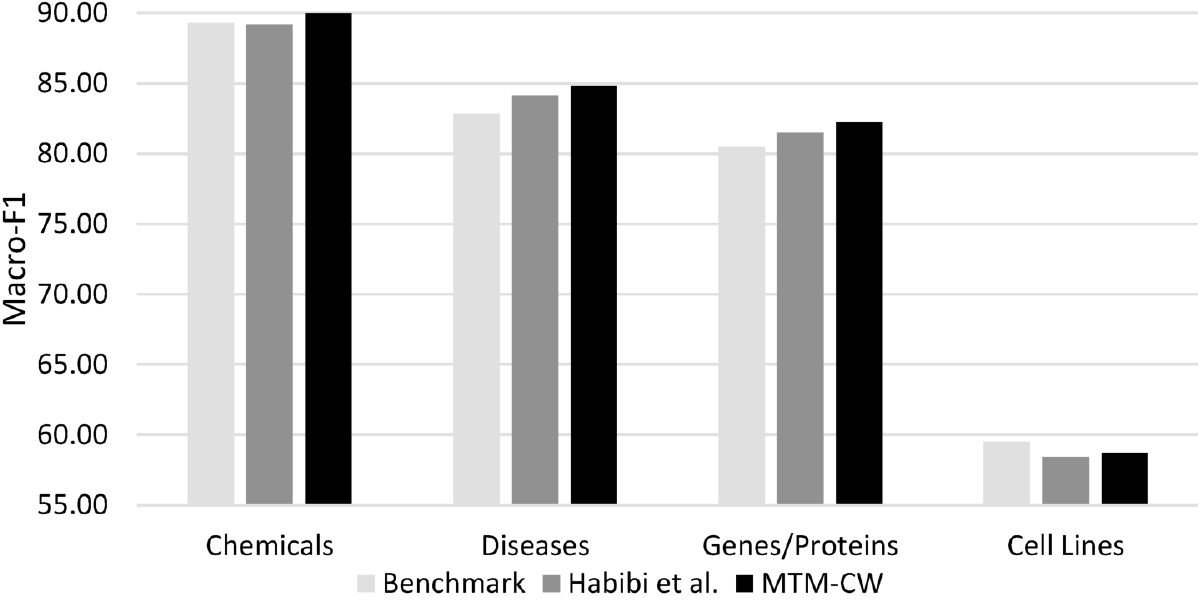}}
			\vspace{-1em}
			\caption{Macro-averaged F1 scores of the proposed multi-task model compared with benchmark on different entities. Benchmark refers to the performance of state-of-the-art BioNER systems.}
			\vspace{-2em}
			\label{bmentity}
		\end{figure}
		
		
		\subsection{Integration of biomedical entity dictionaries}
		A biomedical entity dictionary is a manually-curated list of entity names that belong to a specific entity type. Traditional BioNER systems make heavy use of these dictionaries in addition to other data. To study whether our approach can benefit from the use of entity dictionaries, we retrieve biomedical entity dictionaries for three entity types (i.e., genes/proteins, chemicals and diseases) from the Comparative Toxicogenomics Database (CTD) (\citealp{davis2017comparative}). We use these entity dictionaries in a neural network model in two different ways: (1) dictionary post-processing to match the `O'-labeled entities with the dictionary to reduce the false negative rate, or (2) dictionary feature to provide additional information about words into the word-level BiLSTM. This dictionary feature indicates whether a word sequence consisting of a word and its neighbors is present in a dictionary. We consider word sequences of up to six words, which adds 21 additional dimensions for each entity type.
		We compare the performance of MTM-CW with and without adding dictionaries (Table \ref{dict}).
		
		We observe no significant performance improvement when biomedical entity dictionaries are included into the MTM-CW model at the pre-processing stage. Moreover, including dictionaries at the post-processing stage even hurts the performance. This is presumably due to a higher false positive rate introduced by the dictionaries, when some words share the surface name with dictionary entities but do not share the same meaning or entity types. These results indicate that our multi-task model, by sharing information at both the character and word levels, is able to learn effective data representations and generalize to new data without the use of external lexicon resources.
		
		\begin{table}[]
			\centering
			\caption{F1 scores of the proposed multi-task model using the CTD  entity dictionary. Bold: best scores, *: significantly worse than the MTM-CW model ($p\leq 0.05$), **: significantly worse than the MTM-CW model ($p\leq 0.01$).}
			\label{dict}
			\begin{tabular}{p{2cm}p{1.7cm}p{1.7cm}p{1.7cm}}
				\hline
				Dataset  & MTM-CW   & +Dictionary Feature & +Dictionary Post-process\\
				\hline
				BC2GM    &  \textbf{80.74$\pm$0.04} &  80.70$\pm$0.06 & 61.56$\pm$0.07$^{**}$\\
				BC4CHEMD &  \textbf{89.37$\pm$0.07}  &  88.92$\pm$0.10 & 83.83$\pm$0.09$^{**}$\\
				BC5CDR   & 88.78$\pm$0.12  &  \textbf{88.82$\pm$0.17} & 87.90$\pm$0.06$^{**}$\\
				NCBI-Disease & \textbf{86.14$\pm$0.31}  &  85.48$\pm$0.44 & 83.80$\pm$0.06$^{*}$\\
				JNLPBA   & \textbf{73.52$\pm$0.03}  &  73.35$\pm$0.30 & 63.62$\pm$0.08$^{**}$\\ 
				\hline
			\end{tabular}
			\vspace{-2em}
		\end{table}
		
		\subsection{Comparison on training time}
		All of the neural network models are trained on one GeForce GTX 1080 GPU. We compare the average training time (seconds per sentence) of our method on the 5 main datasets with the baseline neural models in Table \ref{performance}. Since our multi-task model requires training on the 5 datasets together, we calculate and compare the average training time on all datasets instead of on each individual one. We find that our single-task neural model STM is the most efficient among the neural models and almost halves the training time (0.71 s/sent.) when compared to \citeauthor{lample2016neural, habibi2017deep} (1.59 s/sent.). Compared to the single-task model STM, our multi-task model MTM-CW achieves 8.0\% overall F1 improvements with only 5.1\% additional training time. The reason that MTM-CW is slightly slower compared with STM is that it takes a few more epochs for MTM-CW to reach convergence when trained on 5 datasets together.

		\subsection{Case study}
		To investigate the major advantages of the multi-task models compared with the single task models, we examine some sentences with predicted labels (Table \ref{case}). The true labels and the predicted labels of each model are underlined in a sentence.
		
		One major challenge of BioNER is to recognize a long entity with integrity. In Case 1, the true gene entity is ``endo-beta-1,4-glucanase-encoding genes''. The single-task model tends to break this whole entity into two parts separated by a comma, while the multi-task model can detect this gene entity as a whole. This result could due to the co-training of multiple datasets containing long entity training examples. Another challenge is to detect the correct boundaries of biomedical entities. In Case 2, the correct protein entity is ``SMase'' in the phrase ``SMase - sphingomyelin complex structure''. The single-task models recognize the whole phrase as a protein entity. 
		Our multi-task model is able to detect the correct right boundary of the protein entity, probably also due to seeing more examples from other datasets which may contain ``sphingomyelin'' as a non-chemical entity. In Case 3, the adjective words ``human'' and ``complement factor'' in front of ``H deficiency'' should be included as part of the true entity. The single-task models missed the adjective words while the multi-task model is able to detect the correct right boundary of the disease entity. In summary, the multi-task model works better at dealing with two critical challenges for BioNER: (1) recognizing long entities with integrity and (2) detecting the correct left and right boundaries of biomedical entities. Both improvements come from collectively training multiple datasets with different entity types and sharing useful information between datasets.
		
		\begin{table*}[t]
			\small
			\centering
			\caption{Case study of the prediction results from different models. The true labels and the predicted labels of each model are underlined in the sentence. A brief summary of the error type is also included at the end of each example.}
			\label{case}
			\begin{tabular}{llp{14cm}}
				\hline
				\multicolumn{3}{c}{\textbf{Genes/Proteins}} \\
				\hline
				\multirow{4}{*}{Case 1} &True label & This fragment contains two complete \underline{endo - beta - 1, 4 - glucanase - encoding genes}, designated \underline{celCCC} and \underline{celCCG}. \\
				&Habibi & This fragment contains two complete \underline{endo - beta - 1}, \underline{4 - glucanase} - encoding genes, designated \underline{celCCC} and \underline{celCCG}. \\
				&STM & This fragment contains two complete \underline{endo - beta - 1}, \underline{4 - glucanase - encoding genes}, designated \underline{celCCC} and \underline{celCCG}. \\
				&MTM-CW & This fragment contains two complete \underline{endo - beta - 1, 4 - glucanase - encoding genes}, designated \underline{celCCC} and \underline{celCCG}. \\
				Error & \multicolumn{2}{c}{Entity integrity: break a long entity into parts and lose the entity integrity.} \\
				\hline
				\multirow{4}{*}{Case 2} &True label & A model for the  \underline{SMase} - sphingomyelin complex structure was built to investigate how the  \underline{SMase} specifically recognizes its substrate. \\
				&Habibi & A model for the \underline{SMase - sphingomyelin complex structure} was built to investigate how the \underline{SMase} specifically recognizes its substrate. \\
				&STM & A model for the \underline{SMase  - sphingomyelin complex} structure was built to investigate how the \underline{SMase}  specifically recognizes its substrate. \\
				&MTM-CW & A model for the \underline{SMase} - sphingomyelin complex structure was built to investigate how the \underline{SMase}  specifically recognizes its substrate. \\
				Error & \multicolumn{2}{c}{Right boundary error: false detection of non-entity tokens as part of the true entity.} \\
				\hline\hline
				\multicolumn{3}{c}{\textbf{Diseases}} \\
				\hline
				\multirow{4}{*}{Case 3} &True label & ... \underline{human complement factor H deficiency} associated with hemolytic uremic syndrome. \\
				&Habibi & ...human complement factor \underline{H deficiency} associated with hemolytic uremic syndrome. \\
				&STM & ... human \underline{complement factor H deficiency} associated with hemolytic uremic syndrome. \\
				&MTM-CW &... \underline{human complement factor H deficiency} associated with hemolytic uremic syndrome. \\
				Error & \multicolumn{2}{c}{Left boundary error: fail to detect the correct left boundary of the true entity due to some adjective words in front.} \\
				\hline
			\end{tabular}
			\vspace{-2em}
		\end{table*}

	\end{methods}
	
	\vspace{-1em}
	\section{Conclusion}
	We proposed an neural multi-task learning approach for biomedical named entity recognition. The proposed approach, despite being simple and not requiring manual feature engineering, outperformed state-of-the-art systems and several strong neural network models on benchmark BioNER datasets. We also showed through detailed analysis that the strong performance is achieved by the multi-task model with only marginally added training time, and confirmed that the large performance gains of our approach mainly come from sharing character- and word-level information between biomedical entity types.
	
	Lastly, we highlight several future directions to improve the multi-task BioNER model. First, combining single-task and multi-task models might be a fruitful direction. Second, by further resolving the entity boundary and type conflict problem, we could build a unified system for recognizing multiple types of biomedical entities with high performance and efficiency.



	\vspace{-1em}
	\bibliographystyle{natbib}

	\bibliography{document}

\end{document}